\documentclass[10pt,conference]{IEEEtran}
\IEEEoverridecommandlockouts

\usepackage{cite}
\usepackage{amsmath,amssymb,amsfonts}
\usepackage{graphicx}
\usepackage{textcomp}
\usepackage{xcolor}
\def\BibTeX{{\rm B\kern-.05em{\sc i\kern-.025em b}\kern-.08em
    T\kern-.1667em\lower.7ex\hbox{E}\kern-.125emX}}

\usepackage {listings}
\usepackage{tcolorbox}
\usepackage{zref-base}

\makeatletter
\newcounter{mylstlisting}
\newcounter{mylstlines}
\lst@AddToHook{PreSet}{%
  \stepcounter{mylstlisting}%
  \ifnum\mylstlines=1\relax
    \lstset{numbers=none}
  \else
    \lstset{numbers=left}
  \fi
  \setcounter{mylstlines}{0}%
}
\lst@AddToHook{EveryPar}{%
  \stepcounter{mylstlines}%
}
\lst@AddToHook{ExitVars}{%
  \begingroup
    \zref@wrapper@immediate{%
      \zref@setcurrent{default}{\the\value{mylstlines}}%
      \zref@labelbyprops{mylstlines\the\value{mylstlisting}}{default}%
    }%
  \endgroup
}

\newcommand*{\mylstlines}{%
  \zref@extractdefault{mylstlines\the\value{mylstlisting}}{default}{0}%
}
\makeatother

\newcommand{\boldline}[1]{\textbf{\lstinline!#1!}}

\lstdefinelanguage{XML}
{
  morestring=[b]",
  numbersep=5pt,
  morestring=[s]{>}{<},
  morecomment=[s]{<?}{?>},
  stringstyle=\color{black},
  identifierstyle=\color{forestgreen(web)},
  breaklines=true,
  keywordstyle=\color{bostonuniversityred},
  classoffset=1,
  otherkeywords={def, use, dvars, ptrs, cfuncs},
  morekeywords=[1]{parent, module, name, entry, n, p},
  keywordstyle=\color{black}\bfseries,
  classoffset=0,
  tabsize=12
}

\usepackage{algpseudocode}
\usepackage{multirow}
\usepackage{float}
\usepackage{transparent}
\usepackage{marvosym}
\usepackage{calc}
\usepackage{comment}
\usepackage[utf8]{inputenc}
\usepackage[english]{babel}
\usepackage{savesym}
\savesymbol{widering}
\usepackage{yhmath}
\usepackage{tikz}       
\usetikzlibrary{fit}

\usepackage{enumitem}
\usetikzlibrary{arrows.meta}
\usepackage{booktabs}   
\usepackage{siunitx}    
\usepackage{multirow}
\usepackage{wrapfig}
\usepackage{rotating, graphicx}
\usepackage[flushleft]{threeparttable}
\usepackage{tabularx,colortbl}
\usepackage{adjustbox}
\usepackage{url}
\usepackage[linesnumbered,ruled,vlined]{algorithm2e}
\algblockdefx[Foreach]{Foreach}{EndForeach}[1]{\textbf{foreach} #1 \textbf{do}}{\textbf{end foreach}}
\SetKwInput{KwInput}{Input}                
\SetKwInput{KwOutput}{Output}              
\SetKwProg{Fn}{function}{}{}

\usepackage{pdflscape}
\usepackage{soul}
\usepackage{subfigure}
\usepackage{multirow}
\usepackage{wrapfig}
\usepackage{rotating, graphicx}
\usepackage[flushleft]{threeparttable}
\usepackage{tabularx,colortbl}
\usepackage{adjustbox}
\usepackage[linesnumbered,ruled,vlined]{algorithm2e}
\SetKwInput{KwInput}{Input}                
\SetKwInput{KwOutput}{Output}              
\SetKwProg{Fn}{function}{}{}

\usepackage{xcolor}
\definecolor{codegreen}{rgb}{0,0.6,0}
\definecolor{codegray}{rgb}{0.5,0.5,0.5}
\definecolor{codepurple}{rgb}{0.58,0,0.82}
\definecolor{backcolour}{rgb}{0.95,0.95,0.92}
\definecolor{celadon}{rgb}{0.67, 0.88, 0.69}
\definecolor{gray}{rgb}{0.4,0.4,0.4}
\definecolor{darkblue}{rgb}{0.0,0.0,0.6}
\definecolor{cyan}{rgb}{0.0,0.6,0.6}
\definecolor{bostonuniversityred}{rgb}{0.8, 0.0, 0.0}
\definecolor{forestgreen(web)}{rgb}{0.13, 0.55, 0.13}
\definecolor{LightGray}{gray}{0.9}

\usepackage{expl3,xparse}

\ExplSyntaxOn
\NewDocumentCommand \lstcolorlines { O{gray} m }
{
 \clist_if_in:nVT { #2 } { \the\value{lstnumber} }{ \color{#1 } }
}
\ExplSyntaxOff

\makeatletter
\let\old@lstKV@SwitchCases\lstKV@SwitchCases
\def\lstKV@SwitchCases#1#2#3{}
\makeatother
\usepackage{lstlinebgrd}
\makeatletter
\let\lstKV@SwitchCases\old@lstKV@SwitchCases

\lst@Key{numbers}{none}{%
    \def\lst@PlaceNumber{\lst@linebgrd}%
    \lstKV@SwitchCases{#1}%
    {none:\\%
     left:\def\lst@PlaceNumber{\llap{\normalfont
                \lst@numberstyle{\thelstnumber}\kern\lst@numbersep}\lst@linebgrd}\\%
     right:\def\lst@PlaceNumber{\rlap{\normalfont
                \kern\linewidth \kern\lst@numbersep
                \lst@numberstyle{\thelstnumber}}\lst@linebgrd}%
    }{\PackageError{Listings}{Numbers #1 unknown}\@ehc}}
    
\makeatother

\usepackage{hyperref}
\usepackage{amsmath}
\usepackage[nameinlink]{cleveref}
    \crefname{figure}{Figure}{Figures}  
    \crefname{section}{Section}{Sections}   
    \crefname{table}{Table}{Tables}     
    \crefname{listing}{Listing}{Listings} 
    \crefname{algorithm}{Algorithm}{Algorithms} 

\input xy 
\xyoption{all}

\usepackage{amsthm}
\usepackage{ulem}

\theoremstyle{definition}

\theoremstyle{remark}

\usepackage{wasysym}
\usepackage{stmaryrd}

\usepackage{array}

\begin{document}

\title{A Slicing-Based Approach for Detecting and Patching Vulnerable Code Clones}

\author{
    \IEEEauthorblockN{1\textsuperscript{st} Hakam W. Alomari, 
    2\textsuperscript{nd} Christopher Vendome, 
    3\textsuperscript{rd} Himal Gyawali}
    \IEEEauthorblockA{
        \textit{Department of Computer Science and Software Engineering} \\
        \textit{Miami University}, Oxford, Ohio USA \\
        \{alomarhw, vendomcg, gyawalh\}@miamioh.edu
    }
}


\maketitle

\begin{abstract}
Code cloning is a common practice in software development, but it poses significant security risks by propagating vulnerabilities across cloned segments. To address this challenge, we introduce \textsc{srcVul}, a scalable, precise detection approach that combines program slicing with Locality-Sensitive Hashing to identify vulnerable code clones and recommend patches. \textsc{srcVul} builds a database of vulnerability-related slices by analyzing known vulnerable programs and their corresponding patches, indexing each slice’s unique structural characteristics as a vulnerability slicing vector. During clone detection, \textsc{srcVul} efficiently matches slicing vectors from target programs with those in the database, recommending patches upon identifying similarities. Our evaluation of \textsc{srcVul} against three state-of-the-art vulnerable clone detectors demonstrates its accuracy, efficiency, and scalability, achieving 91\% precision and 75\% recall on established vulnerability databases and open-source repositories. These results highlight \textsc{srcVul}'s effectiveness in detecting complex vulnerability patterns across diverse codebases.
\end{abstract}


\section{Introduction}
\label{sec:introduction}
As software systems grow increasingly complex, developers are more relying on code cloning to streamline and accelerate development~\cite{kim2017vuddy, salimi2022vulslicer}. While code clones can be beneficial when used judiciously~\cite{kapser2008cloning, kim2004ethnographic}, they are often seen as suboptimal due to their potential to propagate security vulnerabilities, which increase maintenance costs and reduce software quality~\cite{mayrand1996experiment, baker1995finding, pham2010detection, li2016clorifi}. Vulnerabilities in one instance of code can spread quickly across all cloned segments. Thus, once a vulnerability (e.g., a buffer overflow or use after free) is embedded in a cloned segment, it is likely to persist across multiple locations within or even across projects. This widespread duplication compromises system security, leading to numerous data breaches and network security incidents. A notable example is the OpenSSL Heartbleed vulnerability (CVE-2014-0160)~\cite{Openssl}, which affected a wide range of systems, including websites, OS distributions, and software applications~\cite{kim2017vuddy}. This vulnerability spread rapidly because many systems either used the entire OpenSSL library or cloned segments of it, underscoring the urgent need for robust clone vulnerability detection to safeguard system security.

Each year, numerous vulnerabilities are reported by the National Vulnerability Database (NVD)~\cite{NVD}, Common Vulnerabilities and Exposures (CVE)~\cite{CVE}, and Common Weakness Enumeration (CWE)~\cite{CWE} with over 240,000 CVEs as of October 2024. A significant portion of these vulnerabilities recur due to code cloning and software reuse. Studies across vulnerability databases reveal that recurring vulnerabilities often appear in systems utilizing copied code or common frameworks, where vulnerable code fragments are replicated without prompt application of patches. These reused fragments are structurally similar and retain consistent names for functions and variables, thereby amplifying security risks through cloned, unpatched code~\cite{pham2010detection}. For example, in CVE-2006-3084, 42 systems reused vulnerable code related to privilege escalation, with vulnerabilities identified in 28 of these, including 20 previously unreported cases~\cite{pham2010detection}. The prevalence of code duplication was first highlighted in Baker's 1995 study~\cite{baker1995finding}, which found that 12\% of the X Window System’s 714,479 LOC were duplicated, where some subsystems contain up to 20\% duplicated code. This redundancy complicates maintenance and creates vulnerabilities when bug fixes applied to one instance of copied code are overlooked in others, resulting in unpatched code. 

In this paper, we address the challenge of detecting vulnerable code clones and automatically recommending their patches using variable-level slicing that focuses on vulnerability-related variables ($vr_{vars}$). These variables are identified by comparing the vulnerable code with its patched counterpart, as they are capable of carrying externally controllable data and are particularly susceptible to vulnerabilities when exploited by malicious actions. Essentially, vulnerabilities arise from a combination of insecure syntax on these critical $vr_{vars}$, which play a pivotal role in program behavior and can be direct triggers for vulnerabilities. For example, vulnerabilities such as ``use after free'' occur due to improper memory release associated with a variable~\cite{schuster2015counterfeit}.

Our approach utilizes decompositional program slicing at the variable level, avoiding the limitations of statement-level slicing methods, which often include all dependencies within a statement and capture both vulnerability-related and unrelated variables. In contrast, our method focuses precisely on $vr_{vars}$ within vulnerability-related statements, ensuring thorough analysis by capturing all computations (both forward and backward) directly associated with each variable. This comprehensive slicing strategy not only minimizes irrelevant statements, reducing false positives (FPs), but also preserves the broader vulnerability context, reducing the risk of false negatives (FNs) from overlooking relevant computations. This precision is crucial for vulnerabilities that span non-contiguous code fragments, as vulnerable lines may be distributed across various functions or areas within the codebase.
\section{Background and Motivation}
\label{sec:background}
Code cloning involves creating code components that are identical or similar to existing ones, including various types such as \textit{exact} (Type-1), \textit{renamed} (Type-2), \textit{near-miss} (Type-3), and \textit{semantic} (Type-4) clones~\cite{roy2007survey, alomari2022clone}. Extensive research has been conducted on clone detection~\cite{roy2007survey, bellon2007comparison, rattan2013software, sheneamer2016survey, svajlenko2014evaluating, alomari2020srcclone}, with approaches differing in their \textit{definitions}, \textit{similarity metrics}, and \textit{code analysis depth}. Modern techniques for detecting vulnerable clones, building on traditional methods, continue to encounter significant challenges in both scalability and accuracy~\cite{salimi2022vulslicer, yamaguchi2011vulnerability, woo2022movery, yamaguchi2014modeling, yamaguchi2015automatic, li2016vulpecker, kim2017vuddy, petukhov2008detecting}.

Scalability issues arise as these methods struggle to handle the sheer volume of code in large software repositories, where millions of LOC are common. Thus, many tools have high processing times and require significant computational resources, limiting their effectiveness for real-time or large-scale vulnerability detection tasks. 

Accuracy challenges stem from two main issues. First, many methods tend to return substantial amounts of vulnerability-irrelevant information, as they often operate at granularity levels (e.g., function level) that may not capture the specifics of a vulnerability, leading to over-inclusion and reduced accuracy. Second, detecting near-miss and semantic clones remains difficult because these clones typically involve non-contiguous, reordered, or interwoven statements across different regions of the code. Their structural diversity and distribution throughout the codebase allow them to evade detection by conventional methods, which are often tuned to recognize more straightforward syntactic patterns.

\begin{figure}
\centering
    \includegraphics[width=\columnwidth]{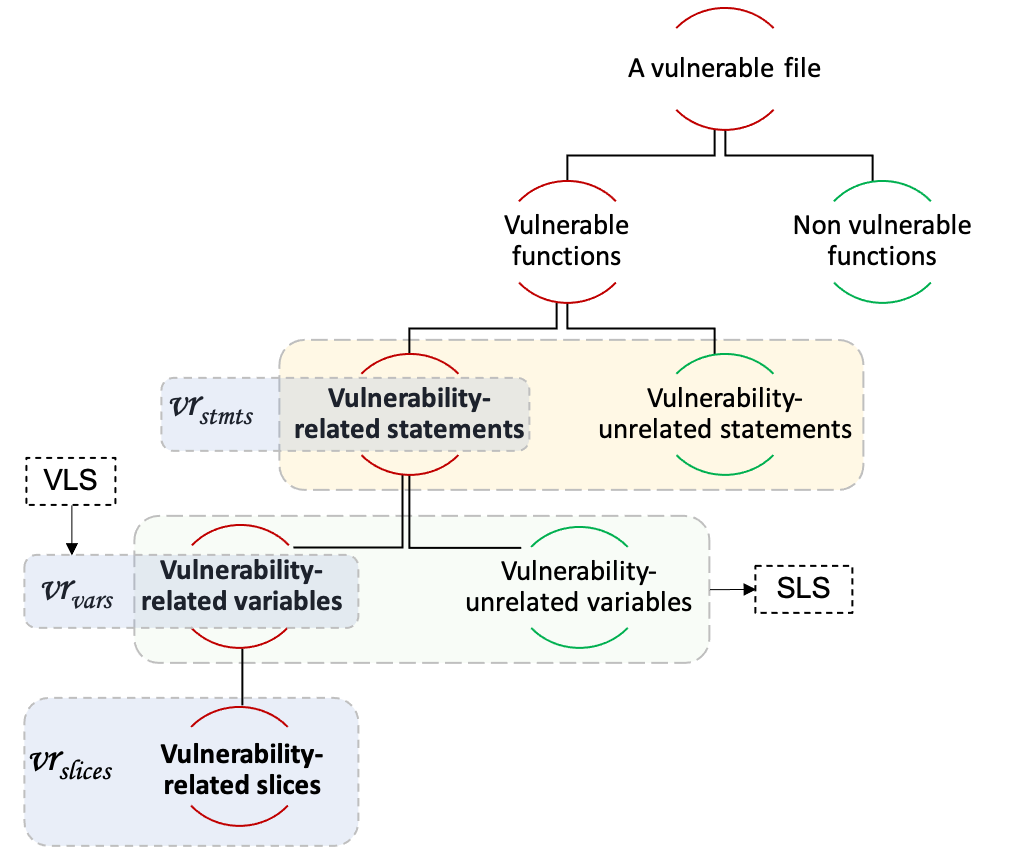}
    \caption{This figure illustrates the comparison between Statement-Level Slicing (SLS) and Variable-Level Slicing (VLS) for identifying vulnerability-related code. The bolded path represents the approach used by \textsc{srcVul}: starting with vulnerability-related statements $vr_{stmts}$, we isolate the vulnerability-related variables $vr_{vars}$ from these statements and then generate vulnerability-related slices $vr_{slices}$. }
    \vspace{-1em}
    \label{fig:archi2}
\end{figure}

Program slicing is a well-established technique for decomposing software into relevant data and control flow dependencies, facilitating an understanding of underlying program semantics~\cite{weiser1984program, tip1994survey}. By tracing these dependencies, slicing identifies both direct and indirect relationships among variables and statements. Various slicing techniques exist~\cite{tip1994survey, silva2012vocabulary, binkley2004survey}, generating different types of slices. However, these methods can struggle with scalability, particularly in large software systems, due to their reliance on Program Dependence Graphs (PDGs)~\cite{ferrante1987program} for slice computation, a computationally intensive process that becomes resource-demanding in larger contexts. 

Recent research has shown that program slicing can improve the accuracy of vulnerable code clone detection by excluding information irrelevant to vulnerabilities~\cite{salimi2022vulslicer, song2020program, xue2018clone}. Additionally, slicing techniques have been applied to enhance general vulnerability detection processes beyond code cloning~\cite{wu2024ultravcs}. Despite its advantages, program slicing remains resource-intensive, though scalable tools like \textsc{srcSlice} have been developed to mitigate these challenges~\cite{alomari2014srcslice, newman2016srcslice}. 


Traditional program slicing considers both control and data dependencies to identify relevant code segments. However, it often operates at the statement level, referred to as Statement-Level Slicing (SLS), where dependencies across entire statements are aggregated. While this approach filters out some vulnerability-unrelated statements within function units, as illustrated in~\cref{fig:archi2}, it may still include irrelevant information when unrelated variables are present within a statement. This aggregation can introduce noise, reducing the precision needed for detecting vulnerabilities tied to specific variables.

To address this limitation, a more fine-grained approach, Variable-Level Slicing (VLS), focuses specifically on the $vr_{vars}$. VLS captures precise data and control dependencies at the variable level, enabling a more targeted analysis of vulnerability-related code segments. By operating at this level of granularity, VLS provides greater accuracy and detail than SLS, improving its utility for detecting vulnerabilities in code.


 
\begin{figure*}
    \begin{minipage}{0.47\textwidth}
\begin{lstlisting}[language=C, firstnumber=696, linebackgroundcolor={\lstcolorlines[pink]{716}}]
static struct snd_info_entry *
snd_info_create_entry(const char *name, struct snd_info_entry *parent,
		      struct module *module)
{
	struct snd_info_entry *entry;
	entry = kzalloc(sizeof(*entry), GFP_KERNEL);
	if (entry == NULL)
		return NULL;
	entry->name = kstrdup(name, GFP_KERNEL);
	if (entry->name == NULL) {
		kfree(entry);
		return NULL;
	}
	entry->mode = S_IFREG | 0444;
	entry->content = SNDRV_INFO_CONTENT_TEXT;
	mutex_init(&entry->access);
	INIT_LIST_HEAD(&entry->children);
	INIT_LIST_HEAD(&entry->list);
	entry->parent = parent;
	entry->module = module;
	if (parent)
		list_add_tail(&entry->list, &parent->children);
	return entry;
}
\end{lstlisting}
\begin{lstlisting}[language=C, numbers=none]
/* Lines 720 - 787 are omitted here */


\end{lstlisting}
\begin{lstlisting}[language=C, firstnumber=788, linebackgroundcolor={\lstcolorlines[pink]{795}}]
		mutex_unlock(&info_mutex);
	}

	/* free all children at first */
	list_for_each_entry_safe(p, n, &entry->children, list)
		snd_info_free_entry(p);

	list_del(&entry->list);
	kfree(entry->name);
	if (entry->private_free)
		entry->private_free(entry);
	kfree(entry);
}
\end{lstlisting}
     \end{minipage}
     \hspace{1.5em}
    \begin{minipage}{0.47\textwidth}
\begin{lstlisting}[language=C, firstnumber=696, linebackgroundcolor={\lstcolorlines[celadon]{716, 717, 719, 720}}]
static struct snd_info_entry *
snd_info_create_entry(const char *name, struct snd_info_entry *parent,
		      struct module *module)
{
	struct snd_info_entry *entry;
	entry = kzalloc(sizeof(*entry), GFP_KERNEL);
	if (entry == NULL)
		return NULL;
	entry->name = kstrdup(name, GFP_KERNEL);
	if (entry->name == NULL) {
		kfree(entry);
		return NULL;
	}
	entry->mode = S_IFREG | 0444;
	entry->content = SNDRV_INFO_CONTENT_TEXT;
	mutex_init(&entry->access);
	INIT_LIST_HEAD(&entry->children);
	INIT_LIST_HEAD(&entry->list);
	entry->parent = parent;
	entry->module = module;
	if (parent) {
		mutex_lock(&parent->access);
		list_add_tail(&entry->list, &parent->children);
		mutex_unlock(&parent->access);
	}
	return entry;
}
\end{lstlisting}
\begin{lstlisting}[language=C, numbers=none]
/* Lines 723 - 797 are omitted here*/
\end{lstlisting}
\begin{lstlisting}[language=C, firstnumber=798, linebackgroundcolor={\lstcolorlines[celadon]{798, 799, 800, 801, 802, 803}}]
	p = entry->parent;
	if (p) {
		mutex_lock(&p->access);
		list_del(&entry->list);
		mutex_unlock(&p->access);
	}
	kfree(entry->name);
	if (entry->private_free)
		entry->private_free(entry);
	kfree(entry);
}
\end{lstlisting}
         \label{fig:motivation(b)}
   \end{minipage}
     \caption{A ``use after free'' vulnerability and its patch in \texttt{linux-5.1.0/sound/core/info.c}, associated with CVE-2019-15214. The left side shows the vulnerable code snippet spanning two functions: \texttt{snd\_info\_create\_entry} (module size = 24 LOC) and \texttt{snd\_info\_free\_entry} (module size = 22 LOC), with highlighted lines representing deleted lines. The right side shows the patched code snippet for the same functions: \texttt{snd\_info\_create\_entry} (module size = 27 LOC) and \texttt{snd\_info\_free\_entry} (module size = 27 LOC), with highlighted lines representing added lines. The line numbers correspond to those in the original vulnerable and patched files.}
     \vspace{-0.5em}
     \label{fig:motivation}
\end{figure*}



To illustrate the significance of our work, we present a code example from Song et al.'s work on detecting vulnerable code clones~\cite{song2020program}. In the code snippet shown in~\cref{fig:motivation} (left), you can observe a critical vulnerability within the Linux kernel's core sound driver, which was identified as CVE-2019-15214. In this specific case, the vulnerability relates to a ``use after free'' scenario occurring in a race condition between disconnection events. This condition could potentially allow a local attacker with the capability to trigger disconnection events, such as hardware removal or addition, to crash the system, corrupt memory, or escalate privileges. The problem arises from the fact that adding and deleting elements in a linked list are performed without proper locking mechanisms. This lack of synchronization becomes problematic when these operations are executed concurrently. To address this vulnerability, the recommended patch, as shown in~\cref{fig:motivation} (right), involves protecting the operations of adding and deleting links with the parent's mutex. 

\begin{figure*}
\centering
    \includegraphics[width=\linewidth]{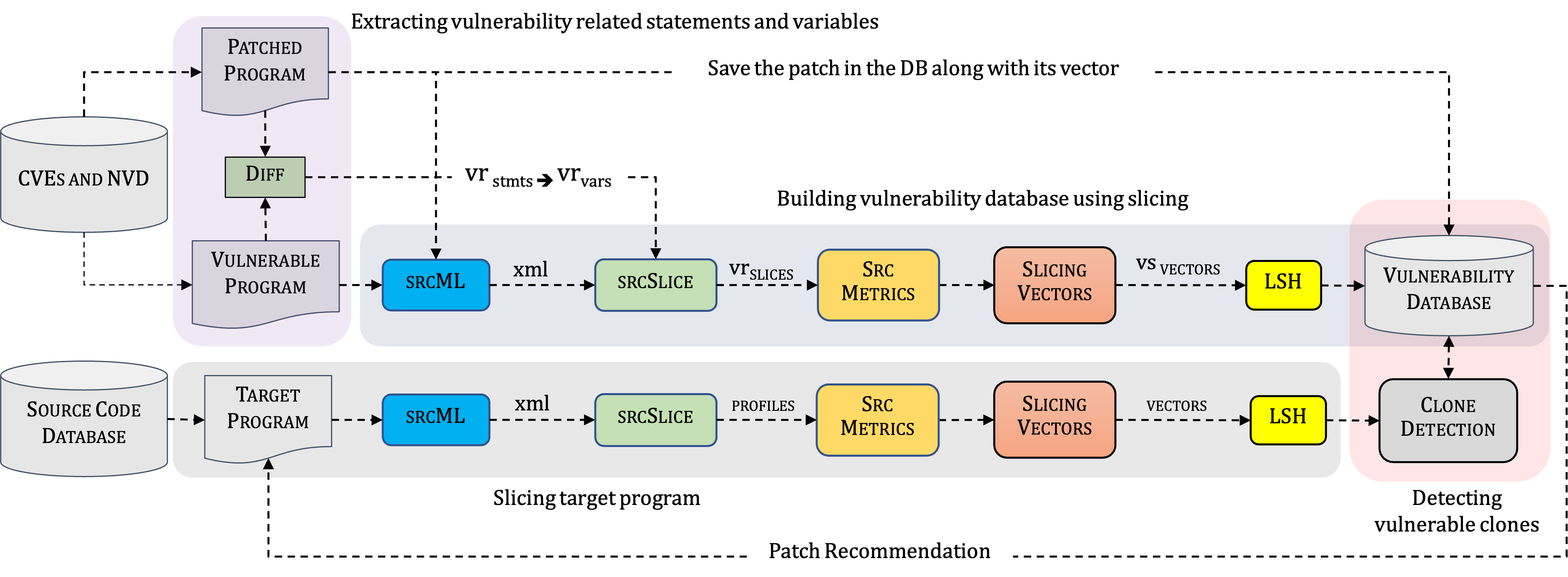}
    \vspace{-2em}
  \caption{Architecture overview of \textsc{srcVul}, illustrating the end-to-end process for detecting vulnerabilities and recommending patches. The workflow involves extracting vulnerability-related code changes, generating slicing-based vectors, and using locality-sensitive hashing for efficient clone detection and patch retrieval from a vulnerability database.} 
    \vspace{-0.5em}
    \label{fig:archi}
\end{figure*}

As illustrated in~\cref{fig:motivation}, this vulnerability affects two separate functions: ``\texttt{snd\_info\_create\_entry}'' and ``\texttt{snd\_info\_free\_entry}.'' In the vulnerable code shown in~\cref{fig:motivation} (left), deleted statements at lines \#716 and \#795 are identified as the $vr_{stmts}$, highlighting only \texttt{parent} and \texttt{entry} as relevant $vr_{vars}$. Reporting both functions as entirely vulnerable would be misleading, as it would include vulnerability-unrelated statements and unrelated variables. For example, \textsc{Vuddy}~\cite{kim2017vuddy}, which detects vulnerabilities at the function level, replaces parameters, variables, data types, and function calls with unified symbols, using function token sequence lengths to filter unrelated code. Functions with matching hashes are marked as vulnerable. Although \textsc{Vuddy} is efficient, it is limited to detecting only minor modifications, such as variable renaming, and struggles with statement insertions or deletions, restricting its scope. Conversely, line-level tools like \textsc{ReDeBug}~\cite{jang2012redebug} focus on cloned code lines, making them suitable primarily for Type-2 clones but less effective in capturing the full context of vulnerabilities that extend across multiple lines or functions.

\section{SrcVul Approach}
\label{sec:approach}
~\Cref{fig:archi} depicts \textsc{srcVul}'s pipeline for both vulnerability detection and patch recommendation by utilizing slicing techniques. By focusing on slice-level granularity, \textsc{srcVul} captures all relevant computations tied to vulnerabilities, ensuring accuracy without including unnecessary code. The system then optimizes slicing vectors for rapid comparison using efficient hashing methods. Additionally, \textsc{srcVul} detects vulnerable code clones, even when code spans non-contiguous sections. 

\subsection{Building the Vulnerability Database}
\label{sec:database}
\subsubsection{Identifying and Extracting $vr_{stmts}$}
The initial stage of \textsc{srcVul} involves analyzing both the vulnerable and patched versions of a program to identify specific program statements associated with the vulnerability, which we term vulnerability-related statements and denote as $vr_{stmts}$. 
To identify these statements, we utilize the Unix utility \textit{diff}, which highlights the added and/or deleted lines between the two versions. This method aligns with established industry practices and is supported by prior research~\cite{salimi2022vulslicer, kim2017vuddy, li2016vulpecker, du2019leopard}.

The identified statements, whether deleted from the vulnerable version or newly added in the patch, are essential for understanding both the root cause of the vulnerability and how the fix addresses it. In other words, the deleted statements include variables directly involved in the vulnerability, providing insight into the issue when we perform slicing on these variables. Conversely, slicing the variables in the added statements enables analysis of the corrected behavior, confirming that the patch resolves the issue and can be applied to similar vulnerable clones. Thus, both deleted and added lines are considered $vr_{stmts}$. The identified $vr_{stmts}$, with differentiation between added and deleted, are stored in XML format using \textsc{sliceDiff}~\cite{alomari2014slice}, an extension of the \textsc{srcML} representation designed to capture code differences and facilitates easy future access. 

\subsubsection{Determining $vr_{vars}$}
After identifying $vr_{stmts}$, all variables within these statements are used as criteria for slicing. These vulnerability-related variables, denoted as $vr_{vars}$ (e.g., \texttt{parent} and \texttt{entry} in~\cref{fig:motivation}), are then utilized to derive the vulnerability-related slices, referred to as $vr_{slices}$.

Studies indicate that many vulnerabilities originate from insecure operations on $vr_{vars}$~\cite{cowan1998stackguard, cowan2000buffer, sui2016svf, szekeres2013sok, schuster2015counterfeit, li2018vuldeepecker}, with system API misuse as a primary cause~\cite{cheng2021deepwukong}. Traditional slicing tools, especially those based on PDGs and SLS, focus on these operations by marking them as ``points of interest'' (POIs) or slicing criteria to identify potential vulnerabilities. Arithmetic operations are also flagged, as they can introduce vulnerabilities through unsafe calculations, making them complementary POIs in vulnerability detection~\cite{wu2024ultravcs, cheng2021deepwukong}. 

In \textsc{srcVul}, the slicing criterion is based on identifying $vr_{vars}$ within $vr_{stmts}$, assuming these variables are central to the vulnerability. This approach aligns with existing literature, which suggests that vulnerabilities often stem from unsafe operations on critical variables like pointers, system APIs, or externally controlled data. By defining $vr_{vars}$ as the slicing criteria, \textsc{srcVul} ensures that slices capture essential information related to specific vulnerability patterns, leading to comprehensive detection across a variety of security threats. 

In the following section, we discuss the resulting slices in more detail and provide a rationale for how the slicing fields of these variables map to various critical types of vulnerabilities identified in the literature, highlighting their role in accurate vulnerability detection.

\begin{table*}
\centering
\caption{Vulnerabilities Addressed and Limitations of \textsc{srcVul}.}
\vspace{-0.5em}
\label{table:srcVul-vulnerability-coverage}
\begin{tabular}{|lclc|}
\toprule
\textbf{Vulnerability Type} & \textbf{Slicing Fields} & \textbf{Description} & \textbf{Reference} \\ 
\midrule

\multicolumn{4}{|l|}{\cellcolor[gray]{.9}\textbf{Handled by \textsc{srcVul}}} \\
\midrule

Memory Management & \textbf{Ptrs, Def, Cfuncs} & Buffer overflow, use-after-free, double free & \cite{cowan1998stackguard} \\

API Misuse & \textbf{Cfuncs, Use} & Unsafe API/system calls (e.g., \texttt{exec}) & \cite{cheng2021deepwukong} \\

Input Handling & \textbf{Use, Cfuncs} & SQL and Command Injection, XSS & \cite{li2018vuldeepecker} \\

Authorization Flaw & \textbf{Dvars, Def, Use} & Controls user roles, tokens for access & \cite{jang2012redebug} \\

Arithmetic \& Logic Errors & \textbf{Use, Dvars} & Logic flaws like integer overflow that bypass security & \cite{szekeres2013sok} \\

Concurrency Management & \textbf{Cfuncs, Ptrs, Dvars} & Detects race conditions and synchronization-related calls & \cite{schuster2015counterfeit} \\

\midrule

\multicolumn{4}{|l|}{\cellcolor[gray]{.9}\textbf{Outside \textsc{srcVul}'s Scope}} \\
\midrule

Semantic Flaws & Not addressed & Logic flaws unrelated to variables & --- \\


Complex Control Flow & Partially (\textbf{Cfuncs}) & Indirect jumps need advanced tools & --- \\

UI Security & Not addressed & Needs UI/UX analysis for phishing, clickjacking & --- \\

\bottomrule
\end{tabular}
\vspace{-1em}
\end{table*}

\subsubsection{Generating $vr_{slices}$}
\label{sec:vrslices}
After establishing the slicing criteria, \textsc{srcVul} extracts $vr_{slices}$, a set of interconnected statements critical to the vulnerability. Using decompositional program slicing, we generate slices for each identified variable. Each slice isolates code sections tied to specific variables, preserving essential data and control dependencies, even if the statements are non-contiguous~\cite{alomari2014srcslice}.

To generate the slices, we employ \textsc{srcSlice}\footnote{~\url{https://www.srcml.org/tools.html}}
\cite{newman2016srcslice}, due to its scalability. 
Since \textsc{srcSlice} leverages the XML representation from~\textsc{srcML}~\cite{collard2011lightweight}, we first utilize \textsc{srcML} on the vulnerable system. Next, \textsc{srcSlice} analyzes this representation to extract information about files, functions, and variables within the system. This information is organized in a three-tiered dictionary, which includes files linked to functions, functions related to variables, and variables associated with slice profiles. The slice profiles generated by \textsc{srcSlice} offer a comprehensive view of each identifier in the system, providing insights into their impact on source code statements through data and control dependencies. Specifically, the slice profiles consist of the following fields:

\begin{itemize}[leftmargin=*]
    \item \texttt{file}, \texttt{function}, and \texttt{variable} names: identify the location of $vr_{vars}$ within the code, tracing vulnerabilities back to specific files and functions.
    
    \item \texttt{def}: lists lines where a variable is defined or redefined, capturing allocations, initializations, and deallocations critical for memory management.
    
    \item \texttt{use}: lists computation lines where the variable’s value is used without modification, aiding in identifying vulnerabilities linked to unvalidated inputs, unsafe data handling, and injection points.
    
    \item \texttt{dvars}: lists variables dependent on the slicing variable’s value, supporting the detection of authorization issues and logic errors where security-sensitive variables affect other program parts.
    
    \item \texttt{ptrs}: lists variables for which the slicing variable acts as a pointer, helping detect memory management vulnerabilities by tracking potential pointer misuse.
    
    \item \texttt{cfuncs}: lists functions called with the slicing variable as an argument, key for spotting API misuse and system vulnerabilities by revealing unsafe function calls involving $vr_{vars}$.
\end{itemize}

\begin{figure}
\begin{lstlisting}[language=xml, firstnumber=1, breaklines=true]
Linux-5.0.10/sound/core/info.c,snd_info_create_entry,module,def{698},use{715},dvars{},ptrs{},cfuncs{}
Linux-5.0.10/sound/core/info.c,snd_info_create_entry,entry,def{700,701,704,709,710,714,715},use{701,702,705,706,711,712,713,717,718},dvars{module,parent},ptrs{},cfuncs{list_add_tail{1},INIT_LIST_HEAD{1},mutex_init{1},kfree{1},kzalloc{2}}
Linux-5.0.10/sound/core/info.c,snd_info_create_entry,parent,def{697},use{714,716,717},dvars{},ptrs{},cfuncs{list_add_tail{2}}
Linux-5.0.10/sound/core/info.c,snd_info_create_entry,name,def{697},use{704},dvars{},ptrs{},cfuncs{kstrdup{1}}
Linux-5.0.10/sound/core/info.c,snd_info_free_entry,entry,def{779},use{782,784,786,790,792,793,795,796},dvars{},ptrs{},cfuncs{private_free{1},kfree{1},list_del{1},snd_info_disconnect{1},list_for_each_entry_safe{3}}
\end{lstlisting}
    \vspace{-0.5em}
    \caption{Slice profiles of the code snippet in~\cref{fig:motivation} (left). Lines \#2 and \#3 represent the slice profiles for the variables \texttt{entry} and \texttt{parent}, respectively, with other irrelevant variables to $vr_{stmt}$ omitted for clarity.}
    \label{fig:vul-slice}
    \vspace{-2em}
\end{figure}

~\cref{fig:vul-slice} illustrates the slice profiles for the code snippet shown in ~\cref{fig:motivation} (left). Each line or entry in the snippet corresponds to a slice profile, structured according to the fields described earlier. After computing these slice profiles, a final pass is executed to account for dependent variables, function calls, and direct pointer aliasing. 
\textsc{srcSlice} utilizes information from function calls and pointer aliases to calculate indirect and inter-procedural slicing details. 

Table~\ref{table:srcVul-vulnerability-coverage} illustrates the mapping between common vulnerability types and the fields within slice profiles, clarifying how each contributes to identifying vulnerability-related attributes. This structured approach enables \textsc{srcVul} to focus on high-risk operations effectively, improving detection accuracy and reducing extraneous information within vulnerability slices.

In ~\cref{fig:motivation}, the variables \texttt{parent} and \texttt{entry} are identified as $vr_{vars}$, so we focus on their slice profiles. \textsc{srcSlice} constructs the complete slice by combining the slice profile of each slicing variable, which includes its \textit{def} and \textit{use} sets, with the profiles of related identifiers found in dependent variables (\textit{dvars}), called functions (\textit{cfuncs}), and pointer (\textit{ptrs}) fields. This combination excludes any lines preceding the variable’s initial definition (i.e., the set {1, ..., $def(v) - 1$}). For instance, as shown in~\cref{fig:vul-slice}, since \texttt{parent} has no dependent variables or pointers, only its use in the call to \texttt{list\_add\_tail()} needs consideration. Consequently, the complete slice for \texttt{parent} can be represented as: \[ def(parent) \cup use(parent) \cup \textit{slice}(list\_add\_tail()) - {1} \] yielding $vr_{slice}(parent) = \{697, 714, 716, 717\}$. Similarly, for the \texttt{entry} variable, only the instance within \texttt{snd\_info\_free\_entry} is treated as a $vr_{vars}$, rather than the one in \texttt{snd\_info\_create\_entry}. This distinction is marked in~\cref{fig:vul-slice} by path names, which specify both file and variable names to differentiate between these instances.

\begin{algorithm}[b]
\caption{Generate $vs_{vectors}$ with Metrics}
\label{alg:GenerateVSVectorsWithMetrics}
\begin{algorithmic}[1]
\Procedure{Generate\_Vulnerability\_Vectors}{$vr_{slices}$, $module\_size$}
    \State Initialize $vs_{vectors}$ as an empty dictionary
    \Foreach{$vr_{slice} \in vr_{slices}$}
        \State Initialize $SC$, $SZ$, $SCvg$, $SI$, $SS$ to 0
        \State $SC \gets \frac{|vr_{slice}|}{module\_size}$ 
        \State $SZ \gets \sum \text{(statement count of each } vr_{slice})$
        \State $SCvg \gets \frac{SZ}{module\_size}$ 
        \State $SI \gets \frac{\text{count of unique identifiers in }\{dvars \cup ptrs \cup cfuncs\}}{module\_size}$
        \State $S_f \gets \text{first statement line in } vr_{slice}$
        \State $S_l \gets \text{last statement line in } vr_{slice}$
        \State $SS \gets \frac{S_l - S_f}{module\_size}$ 
        \State $vs\_vector \gets \langle SC, SCvg, SI, SS \rangle$ 
        \State $vs_{vectors}[vr_{slice}] \gets vs\_vector$
    \EndForeach
    \State \textbf{return} $vs_{vectors}$
\EndProcedure
\end{algorithmic}
\end{algorithm}
\vspace{-0.5em}

\subsubsection{Computing Slice-based Metrics}
\label{sec:metrics}
\vspace{0.2cm}
Next, \textsc{srcVul} computes a range of slice-based cognitive complexity metrics using the information from the $vr_{slices}$. 
These metrics provide valuable insights into the code slices and are key to identifying potential vulnerabilities. 

For each $vr_{slice}$, \textsc{srcVul} computes the following metrics:

\begin{itemize}  [leftmargin=*]
    \item \textit{Slice Count} ($SC$): This metric quantifies the number of slices in relation to the module size. It indicates the number of slice profiles combined to form the final slice. 
    
    \item \textit{Slice Size} ($SZ$): This metric represents the total number of statements within a complete final slice. It plays a role in the calculation of the $SCvg$ metric. 
    
    \item \textit{Slice Coverage} ($SCvg$): This metric measures the slice size relative to the module size, quantified in LOC.
    
    \item \textit{Slice Identifier} ($SI$): This metric counts the unique identifiers, including variables and method invocations, contained within a slice relative to the module's size. They are drawn from the \textit{dvars}, \textit{ptrs}, and \textit{cfuncs} fields within the slice. 
    
    \item \textit{Slice Spatial} ($SS$): This metric represents the spatial distance, measured in LOC, between the initial definition and the final use of the slicing variable, relative to the module size. 
\end{itemize}

These metrics were originally introduced by Alqadi et al.~\cite{alqadi2019relationship} and have been adapted with specific modifications for vulnerability detection. Notably, Slice Count (SC) emphasizes constructs like variables, pointers, and function calls instead of all statements. Similarly, Slice Identifier (SI) extends Alqadi's general definition by explicitly focusing on unique identifiers, such as variables, pointers, and function calls, which are particularly relevant in analyzing vulnerabilities. Additionally, Slice Spatial (SS) focuses on variable definition and usage to better highlight patterns associated with vulnerabilities. These modifications make the metrics more effective for identifying vulnerabilities while building on Alqadi’s foundational work.

~\cref{alg:GenerateVSVectorsWithMetrics} depicts the algorithm \textsc{srcVul} uses to calculates these metrics. We integrate these metrics into our methodology due to their demonstrated effectiveness~\cite{alqadi2020slice}, in enhancing our understanding of program behavior compared to traditional code-based metrics. Also, these metrics play a pivotal role in the clone detection process, and the metrics are subsequently used to generate slicing vectors for the system's slices. 

\subsubsection{Encoding and Archiving $vr_{slices}$}
\label{sec:vectors}
After computing the slice-based metrics for each $vr_{slice}$, \textsc{srcVul} proceeds to encode these slices into a set of vulnerability slicing vectors, denoted by $vs_{vectors}$ using~\cref{alg:GenerateVSVectorsWithMetrics}. These vectors abstract essential characteristics and cognitive complexity metrics into a format that is conducive to analysis, comparison, and efficient storage. The use of $vs_{vectors}$ enables \textsc{srcVul} to effectively compare slices from known vulnerabilities with slices from target code during the vulnerability detection process. Each $vs_{vector}$ is represented as a single numerical vector with fixed dimensions, where each dimension corresponds to a specific metric of the $vr_{slice}$. A possible representation of the $vs_{vector}$ is as follows:\[ vs_{vector}(v) = \langle SC(v), \; SCvg(v), \; SI(v), \; SS(v) \rangle\]

Consider our motivating example in~\cref{fig:motivation} (left) and the corresponding $vr_{slices}$ shown in~\cref{fig:vul-slice}. Next, we present the calculations of slice-based metrics for one variable, \texttt{parent}, and its $vr_{vectors}$. Given that the module spans from line 696 to line 719, the module size is 24 LOC. The calculated metrics for \texttt{parent} are as follows: 

\begin{itemize} [leftmargin=*]
    \item \textbf{Slice Count} \((SC)\): For \texttt{parent}, with only one slice profile: \[SC = \frac{\text{\# Slice Profiles}}{\text{Module Size}} = \frac{1}{24} \approx 0.042\]
    
    \item \textbf{Slice Size} \((SZ)\): Based on the final slice for \texttt{parent} with lines \{697, 714, 716, 717\}, we have: \[ SZ = \text{\# Statements in Complete Final Slice} = 4 \]

    \item \textbf{Slice Coverage} \((SCvg)\): This metric represents the ratio of the slice size to the module size: \[ SCvg = \frac{SZ}{\text{Module Size}} = \frac{4}{24} \approx 0.167 \]

    \item \textbf{Slice Identifier} \((SI)\): For \texttt{parent}, there is one unique identifier, \texttt{list\_add\_tail}, so: \[ SI = \frac{\text{\# Unique Identifies in Slice}}{\text{Module Size}} = \frac{1}{24} \approx 0.042 \]

    \item \textbf{Slice Spatial} \((SS)\): For \texttt{parent}, the first statement is at line 697, and the last statement is at line 717: \[ SS = \frac{S_l - S_f}{\text{Module Size}} = \frac{717 - 697}{24} = \frac{20}{24} \approx 0.833 \]
\end{itemize}

Using the computed metrics,  \[ vs_{vector}(\texttt{parent}) = \langle 0.042, 0.167, 0.042, 0.833 \rangle \] 

After this process, we've transformed the $vr_{slices}$ into a collection of $vs_{vectors}$ and stored them in a database. We would perform a similar analysis on our target system to generate vectors and compare them against those in the database to identify potential vulnerabilities and recommend patches.

\subsection{Slicing the Target Program \& Detecting Vulnerable Clones}
\label{sec:slicing-target}

\subsubsection{Slicing Target Program}
For vulnerability detection, \textsc{srcVul} generates slicing vectors for the \textit{target program} (i.e., the program with potential security vulnerabilities) through a process similar to that used for building the database. \textsc{srcVul} leverages \textsc{srcSlice} to consider all available variables within the codebase, ensuring a comprehensive analysis. This process generates program slices that encompass all variables in the code, providing broad coverage of potential vulnerability-related dependencies. Similar to the database construction phase, these slices are processed to compute slice-based metrics and are subsequently transformed into vectors similar to $vs_{vectors}$. The resultant slicing vectors are then used during the clone detection process.

While \textsc{srcVul} includes all variables in the analysis, the similarity-based clone detection mechanism ensures that only slices with relevant patterns are matched. Slice-based metrics abstract the behavior and roles of variables, allowing the approach to identify relevant slices even when only a subset of function variables is involved in a vulnerability. This mechanism mitigates the impact of unrelated variables on the detection process.

As shown in the evaluation section, such situations do not significantly impede the effectiveness of the approach. \textsc{srcVul} demonstrates its capability to identify relevant slices despite the inclusion of all variables, providing robustness while ensuring that potential vulnerabilities tied to less obvious dependencies are not overlooked.

\begin{figure}
\begin{lstlisting}[language=C, firstnumber=707]
static struct snd_info_entry *
snd_info_create_entry(const char *name, struct snd_info_entry *parent)
{
	struct snd_info_entry *entry;
	entry = kzalloc(sizeof(*entry), GFP_KERNEL);
	if (entry == NULL)
		return NULL;
	entry->name = kstrdup(name, GFP_KERNEL);
	if (entry->name == NULL) {
		kfree(entry);
		return NULL;
	}
	entry->mode = S_IFREG | S_IRUGO;
	entry->content = SNDRV_INFO_CONTENT_TEXT;
	mutex_init(&entry->access);
	INIT_LIST_HEAD(&entry->children);
	INIT_LIST_HEAD(&entry->list);
	entry->parent = parent;
	if (parent)
		list_add_tail(&entry->list, &parent->children);
	return entry;
}
\end{lstlisting}
         \vspace{-0.7em}
         \label{fig:target(b)}
     \caption{Target code snippet for the function ``\texttt{snd\_info\_create\_entry}''. 
    This snippet is a clone of the vulnerable code shown in~\cref{fig:motivation} (left), with a module size of 22 LOC.}

     \label{fig:target}
     \vspace{-1em}
\end{figure}

\begin{figure}
\begin{lstlisting}[language=xml, firstnumber=1]
Linux-4.14.76/sound/core/info.c, snd_info_create_entry, entry, def{710, 711, 714, 719, 720, 724}, use{711, 712, 715, 716, 721, 722, 723, 726, 727}, dvars{parent}, pointers{}, cfuncs{kzalloc{2}, kstrdup{1}, kfree{1}, mutex_init{1}, INIT_LIST_HEAD{2}, list_add_tail{1}}
Linux-4.14.76/sound/core/info.c, snd_info_create_entry, parent, def{708}, use{724, 725, 726}, dvars{}, pointers{}, cfuncs{list_add_tail{2}}
Linux-4.14.76/sound/core/info.c, snd_info_create_entry, name, def{708}, use{714}, dvars{}, pointers{}, cfuncs{kstrdup{1}}
\end{lstlisting}
    \vspace{-0.5em}
    \caption{Slice profiles for target code snippet in~\cref{fig:target}.}
    \label{fig:target-slice}
     \vspace{-1em}
\end{figure}

To illustrate this process, we've selected Linux kernel version 4.14.76 with approximately 17 MLOC as our target program for analysis. To streamline our presentation, we provide a code snippet in~\cref{fig:target}, which corresponds to a specific function identified by \textsc{srcVul} within this version of the kernel. 
This function is recognized as a clone of the vulnerable motivation example from Linux kernel version 5.1.0, in~\cref{fig:motivation} (left). 

While both functions serve the same purpose, there are subtle distinctions. In~\cref{fig:motivation} (left), an additional argument, \texttt{module}, is introduced in the function \texttt{snd\_info\_create\_entry} signature (line 698). Also, line 709, \texttt{entry->mode} is set to \texttt{S\_IFREG | 0444}, indicating read permissions for a regular file. In~\cref{fig:target}, \texttt{entry->mode} is set to \texttt{S\_IFREG | S\_IRUGO}, essentially conveying the same meaning as read-only access for a regular file. Finally, in~\cref{fig:motivation} (left), the line \texttt{entry->module = module;} (line 715) is specific to the code in~\cref{fig:motivation} (left) and is absent in~\cref{fig:target}.

The slice profiles for the target code in~\cref{fig:target} are presented in~\cref{fig:target-slice}. Following the same process, the final slice-based metrics and vector for \texttt{parent} are calculated as: \[ \langle SC, SCvg, SI, SS \rangle = \langle 0.045, 0.182, 0.045, 0.818 \rangle \]

\subsubsection{Detecting Vulnerable Clones}
To detect vulnerable clones, the similarities between slicing vectors in the database and those generated from the target program must be calculated. However, given the potentially large number of $vs_{vectors}$ in the database and the size of the target software system, performing pairwise similarity analysis is computationally infeasible. To address this challenge, we implemented Locality Sensitive Hashing (LSH)~\cite{indyk1998approximate}, which has been used in prior clone detection approaches ~\cite{alomari2022clone, deckard}. Our implementation followed the methodology employed by the \textsc{srcClone} clone detector for handling slice clones~\cite{alomari2022clone}. LSH hashes vectors into buckets, grouping items that are likely to match, thereby significantly reducing the number of required comparisons.

The LSH implementation in \textsc{srcVul} uses cosine similarity, a measure well-suited for capturing structural similarities in slicing vectors~\cite{salton1975vector}, with a predefined similarity threshold of 0.8. This threshold means that two slicing vectors must share at least 80\% similarity in their MinHash signatures to be considered part of the same clone group or pair. The choice of 0.8 was informed by preliminary experiments, where thresholds ranging from 0.6 to 0.9 were tested. Lower thresholds (e.g., 0.6) improved recall but increased false positives, while higher thresholds (e.g., 0.9) enhanced precision but missed semantically similar vectors. The threshold of 0.8 struck an optimal balance, achieving high precision while maintaining effective vulnerability detection.

Thus, \textsc{srcVul} minimizes the number of direct pairwise comparisons required by grouping similar vectors together based on this threshold, ensuring both computational efficiency and accurate vulnerability detection.

To demonstrate this process, consider our motivating example in~\cref{fig:motivation} (left) along with the vulnerable code clone identified in~\cref{fig:target}. Focusing on the variable of interest, \texttt{parent}, we determined that its slicing vector is \( vs_{vector}(\texttt{parent}) = \langle 0.042, \; 0.167, \; 0.042, \; 0.833 \rangle \). Now, when examining the slicing vector of \texttt{parent} in the vulnerable code clone from~\cref{fig:target}, we find that \( vs_{vector}(\texttt{parent}) = \langle 0.045, \; 0.182, \; 0.045, \; 0.818 \rangle \). Using Cosine similarity, we calculate a similarity between \( v_1 \) and \( v_2 \) of approximately 0.9994, indicating a very high similarity. This similarity demonstrates that the two vectors are nearly identical in direction, supporting the notion that the two code segments likely represent clones with similar vulnerability characteristics. With such a high similarity, the LSH process effectively hashes these vectors into the same or nearby buckets, enabling \textsc{srcVul} to quickly identify and compare them without requiring exhaustive pairwise comparisons.

\begin{algorithm} [t]
\caption{Detect Vulnerable Clones and Patches}
\label{alg:DetectVulnerableClones}
\begin{algorithmic}[1]
\Procedure{Detect\_Clones\_With\_Patch}{$target\_program$, $vul\_database$, $LSH\_index$, $similarity\_threshold$}
    \State Initialize $detected\_vul\_clones$ as an empty list
    \State $target\_slices \gets \text{srcSlice}(target\_program)$
    
    \State $vs\_vectors \gets \text{Generate\_Vectors}(target\_slices, \text{module\_size})$ 
    
    \Foreach{$(target\_slice, vs\_vector) \in vs\_vectors$}
        \State $similar\_vectors \gets \text{LSH\_index.Query}(vs\_vector, similarity\_threshold)$
        
        \Foreach{$db\_vector \in similar\_vectors$}
            \State Retrieve $db\_slice$ and $patch$ from $vulnerability\_database$ corresponding to $db\_vector$
            \State Calculate $similarity \gets \text{Cosine\_Similarity}(vs\_vector, db\_vector)$
            \If{$similarity \ge similarity\_threshold$}
                \State Add $(target\_slice, db\_slice, patch)$ to $detected\_vulnerable\_clones$
                \State Recommend $patch$ for $target\_slice$ as a fix
        \EndForeach
    \EndForeach
    
    \State \textbf{return} $detected\_vulnerable\_clones$
\EndProcedure
\end{algorithmic}
\end{algorithm}
\vspace{-0.5em}

\subsection{Patch Recommendation}
\label{sec:patch}

Previous work on vulnerability detection lacks integrated patch recommendation capabilities~\cite{jang2012redebug, kim2017vuddy, salimi2022vulslicer, li2016vulpecker, li2018vuldeepecker, wu2024ultravcs}. In contrast, \textsc{srcVul} combines vulnerability detection with patch recommendation, enabling a streamlined approach to remediating vulnerabilities. By associating each ($vs_{vector}$) with a specific patch, \textsc{srcVul} not only identifies vulnerable clones but also provides an efficient patch retrieval.

Each patch in \textsc{srcVul} is indexed by its associated slicing vector, which serves as a unique identifier. The vector is derived from the vulnerability-related slices ($vr_{slices}$), representing the structural and semantic characteristics of vulnerable code fragments. For instance, if a vulnerable program's slicing vector is $\langle 0.042, 0.167, 0.042, 0.833 \rangle$, this vector is stored in the database as a lookup key, mapping directly to the associated patch. Because the vulnerabilities are mapped to the patches, \textsc{srcVul} is able to retrieve the patches after performing the vulnerability clone detection. 

\cref{alg:DetectVulnerableClones} summarizes this process of detecting vulnerable clones and recommending patches, where the unique slicing vectors streamline the search and retrieval of specific patches. This integrated approach in \textsc{srcVul} provides a practical solution to vulnerability detection and remediation.

\section{Experimental Methodology}
\label{sec:methodology}
For vulnerability detection, we assess \textsc{srcVul} across three key dimensions: accuracy (precision and recall), performance (execution time), and scalability. Additionally, we evaluate \textsc{srcVul}'s capability to recommend patches for detected vulnerabilities. To provide a thorough comparative analysis, \textsc{srcVul} is benchmarked against other leading vulnerable clone detectors. For slice-based approaches, we evaluated against \textsc{VulSlicer}~\cite{salimi2022vulslicer}; relies on PDGs for slicing source code and is designed for vulnerable code clone detection. We also evaluated against non-slice-based approaches, \textsc{Vuddy}~\cite{kim2017vuddy} and \textsc{ReDeBug}\cite{jang2012redebug}, for a comprehensive assessment.

To systematically investigate \textsc{srcVul}'s effectiveness, we address the following research questions:

\begin{itemize}[leftmargin=*] 
    \item \textbf{RQ1}: How accurately does \textsc{srcVul} identify vulnerable code clones compared to other detection methods? 
    \item \textbf{RQ2}: How does \textsc{srcVul} perform in terms of scalability in comparison with existing approaches?
    \item \textbf{RQ3}: How effective is \textsc{srcVul} in recommending appropriate patches for identified vulnerabilities? 
\end{itemize}

\textbf{RQ1} investigates whether the vulnerable code clones detected by \textsc{srcVul} are comparable in accuracy and scope to those identified by other methods. \textbf{RQ2} investigate the \textsc{srcVul}'s scalability and efficiency compared to existing approaches. \textbf{RQ3} evaluates the effectiveness of \textsc{srcVul}'s patch recommendation, specifically in its ability to suggest suitable fixes for detected vulnerabilities.

\textbf{Replication Package:} The replication package can be found at:  \url{https://github.com/alomarhw/srcVul}. It includes the \textsc{srcVul} system for detecting and patching vulnerable code clones, organized into modules for slicing, clone detection, and datasets. Detailed instructions for generating slicing vectors, analyzing vulnerabilities, and reproducing results are provided. The evaluation was performed on a computer with a 3.2 GHz 8-Core Intel processor, 32 GB of DDR4 RAM, and running MacOS.

\subsection{Data Collection}
In order to construct a comprehensive and representative database for our assessment, we sourced data from the National Vulnerability Database (NVD)~\cite{NVD}, GitHub, \textsc{VulSlicer}~\cite{VulSlicer-DB} and \textsc{Vuddy}~\cite{Vuddy-DB}, selecting relevant vulnerabilities from publicly disclosed Common Vulnerabilities and Exposures (CVEs)~\cite{CVE, CVEDetails}.  For each CVE, we gathered associated source code files and patch files containing \texttt{diff} data, indicating code changes. These files were gathered by systematically accessing the reference links provided within each CVE, which map to source code and patch information. We utilized a web crawler to navigate through reference links and locate pages that provide code diffs or patch information. Each CVE entry in our dataset includes the following metadata: i) the CVE ID and description to contextualize the vulnerability type, ii) the vulnerable code and corresponding patch, and iii) versioning details to align the data with specific project versions.

From this collection process, we collected a total of 65,195 \texttt{diff} files. Of these, 60,145 files contain only C and C++ code as either added or deleted lines (excluding comments). Specifically, 13,553 files contain added code lines only, 714 files contain only deleted lines, and 45,878 files contain both added and deleted lines. Overall, these \texttt{diff} files represent 3,666 unique CVEs across 512 projects and sub-projects, with multiple \texttt{diff} files per CVE where necessary  (e.g., for vulnerabilities affecting multiple files or project versions). 

In total, 1,145,200 LOC were deleted in the patches, and 2,869,167 LOC were added. This resulted in a total of 2,350,306 $vr_{stmts}$. This number is not equal to the summation of the added and deleted lines, as lines deleted and then re-added are only counted once. From the $vr_{stmts}$, we found a total of 2,887,118 $vr_{vars}$. 

~\cref{table:srcVul-vulnerability-coverage} depicts the categories of vulnerabilities covered by \textsc{srcVul}. We grouped the data by analyzing comments and descriptive content within each CVE's \texttt{diff} file, using keywords specific to well-known vulnerabilities. The categorization resulted in the following vulnerability mapping: 2,335 \textit{Memory Management}, 686 \textit{Authorization Flaws}, 1,085 \textit{Concurrency Management}, 211 \textit{Input Handling}, 13,470 \textit{API Misuse}, 992 \textit{Logic Errors}, and 41,366 \textit{Uncategorized} (these include vulnerabilities outside the defined categories, which may require further review for precise classification). 
These 18,779 \texttt{diff} files correspond to 1,438 unique CVEs associated with the selected target programs, as discussed in the next section. The slices for the $vr_{vars}$ associated with each CVE are then computed and $vs_{vectors}$ are generated as described in~\cref{sec:approach}. These vectors are associated with CVE information (e.g., CVE ID, original and patched code, commit hash, etc.) 

\begin{table*}
\centering
\caption{Accuracy of \textsc{srcVul} in comparison to other techniques. \#CVEs = number of CVEs found in the vulnerability database. P = Precision, R = Recall., T = Time in seconds. LOC = code counted using cloc no comments nor blank. }
\vspace{-0.5em}
\label{tab:q1}
\begin{tabular}{lcccccccccccccc}
\toprule
\multirow{2}{*}{\textbf{Target}} & \multirow{2}{*}{\textbf{Files}} & \multirow{2}{*}{\textbf{\#CVEs}} & \multirow{2}{*}{\textbf{LOC}}& \multicolumn{5}{c}{\textsc{srcVul}} & \multicolumn{2}{c}{\textsc{VulSlicer}$^+$} & \multicolumn{2}{c}{\textsc{ReDeBug}$^-$} & \multicolumn{2}{c}{\textsc{Vuddy}$^*$} \\
 \cmidrule(lr){5-9} \cmidrule(lr){10-11} \cmidrule(lr){12-13} \cmidrule(lr){14-15}
 &  & & & \textbf{TPs} & \textbf{FNs} & \textbf{FPs} & \textbf{P.} & \textbf{R.} & \textbf{P.} & \textbf{R.} & \textbf{P.} & \textbf{R.} &   \textbf{P.} & \textbf{R.}  \\
\midrule
\texttt{kernel} & 54,745 & 1311 & 17,373,016 & 1000 & 311 & 90 & 0.92 & 0.76 & 0.95 & 0.74 &  0.56 & 0.62 &  0.65 & 0.56  \\
\texttt{samba} & 6,759 & 71 & 2,622,156 & 45 & 26 & 8 & 0.85 & 0.63 &  0.88 & 0.37 &  0.74 & 0.32 &  0.63 & 0.55  \\
\texttt{libvirt} & 2,814 & 43 & 2,101,222 & 30 & 13 & 5 & 0.86 & 0.70 & 0.46 & 0.14 &  0.78 & 0.67 &  0.65 & 0.60  \\
\texttt{libgd} & 530 & 13 & 106,498 & 7 & 6 & 3 & 0.70 & 0.54 & 1.00 & 0.15 &   0.63 & 0.40 &  0.73 & 0.33  \\
\midrule
\textbf{Total} & 64,848 & 1438 & 22,202,892 & 1082 & 356 & 106 & 0.91 & 0.75 & 0.91 & 0.56 &   0.65 & 0.38 &   0.79 & 0.42  \\
\bottomrule
\end{tabular}
\vspace{-0.5em}
\end{table*}

\subsection{Design of Experiments}
\label{sec:design}
To evaluate the effectiveness (\textbf{RQ1}), we evaluated \textsc{srcVul} using the same open-source systems that were analyzed in prior work for the tools that we compared against, ensuring a fair and consistent comparison. 
From \textsc{VulSlicer}, \textsc{ReDeBug}, and \textsc{Vuddy}, we evaluated our approach on the \texttt{Linux kernel}, \texttt{libgd}, \texttt{Libvirt},  and \texttt{Samba}.

To evaluate the accuracy of \textsc{srcVul}, we measured the number of true positives (TPs), false positives (FPs),  false negatives (FNs), precision, recall, and F1-score with respect to the detection of vulnerable clones within a controlled set of target programs that are not presented in our known vulnerability database. This approach ensures that each detection made by \textsc{srcVul} can be directly compared against established vulnerabilities, providing an unbiased and reliable measure of detection accuracy. We selected the following target programs for testing: \texttt{libvirt-1.1.0}, 
\texttt{linux-4.14.76}, \texttt{libgd-2.3.0}, and \texttt{samba-4.0.26}. The number of CVEs from the 1,438 identified in our database for each of these systems is provided in~\cref{tab:q1}. 


\textsc{srcVul} was run on each target program to generate the $vs_{vectors}$ and identify clone candidates corresponding to known vulnerability vectors in the database. The ground truth for vulnerabilities in the target systems was established by comparing the detected vulnerabilities against entries in the \textsc{srcVul} database, which was built from known vulnerabilities in the CVE/NVD datasets. Matches between $vs_{vectors}$ in the target systems and the database entries were treated as true positives (TPs). If a detected clone did not correspond to any CVE in the database, it was classified as a false positive (FP). Conversely, if a known vulnerability from the database was not detected in the target system, it was considered a false negative (FN). For cases where the similarity between a detected clone and an entry in the database fell within the predefined similarity threshold (e.g., 80\%), manual verification was conducted. This manual process involved evaluating whether the detected slice corresponded to the same vulnerability described in the database entry, ensuring accurate classification of true and false positives.

Other state-of-the-art approaches rely on their own matching criteria (e.g., function signatures or line-level granularity) for evaluation. Ground truth vulnerabilities for these approaches are derived from their respective datasets or the CVE/NVD repositories. Metrics such as true positives (TP), false positives (FP), and false negatives (FN) are computed similarly to \textsc{srcVul}, based on whether the detected vulnerabilities align with the known vulnerabilities in the respective ground truth datasets. This consistent methodology ensures a fair comparison of detection accuracy across all approaches.

\begin{figure}
\centering
    \includegraphics[width=\columnwidth]{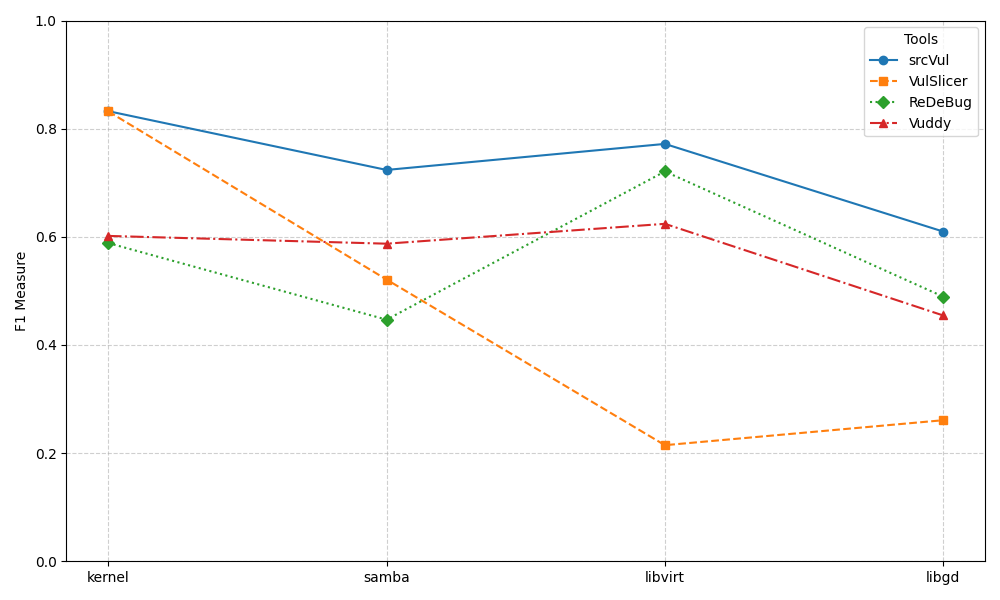}
    \vspace{-2em}
    \caption{F1 Measure values are shown for \textsc{srcVul} and other detectors across target systems. }
    \label{fig:archi3}
\end{figure}

A notable threat to validity in this evaluation is the potential lack of representativeness of the selected target programs relative to the broader set of CVEs in our database. This could mean that some FNs might arise not because \textsc{srcVul} failed to detect a clone, but rather because the specific versions of the target programs may not exhibit characteristics or code patterns associated with other CVEs for that system within the database. With these counts, we calculated precision, recall, and the F1-score (the harmonic mean of precision and recall) for each target system. The comprehensive results of \textsc{srcVul}, along with comparisons to other clone detection tools, are presented in~\cref{tab:q1} and~\cref{fig:archi3}. As shown, the precision and recall metrics highlight \textsc{srcVul} as the most effective tool, with a high precision of 91\% and recall of 75\%, indicating strong accuracy and coverage in detecting vulnerabilities. In contrast, while \textsc{VulSlicer} achieves similar precision (91\%), its recall is lower at 56\%, suggesting it may miss more vulnerabilities. \textsc{ReDeBug} and \textsc{Vuddy} have lower precision and recall scores, with \textsc{ReDeBug} at 65\% precision and 38\% recall, and \textsc{Vuddy} at 79\% precision but only 42\% recall.

\begin{figure}
\centering
    \includegraphics[width=\columnwidth]{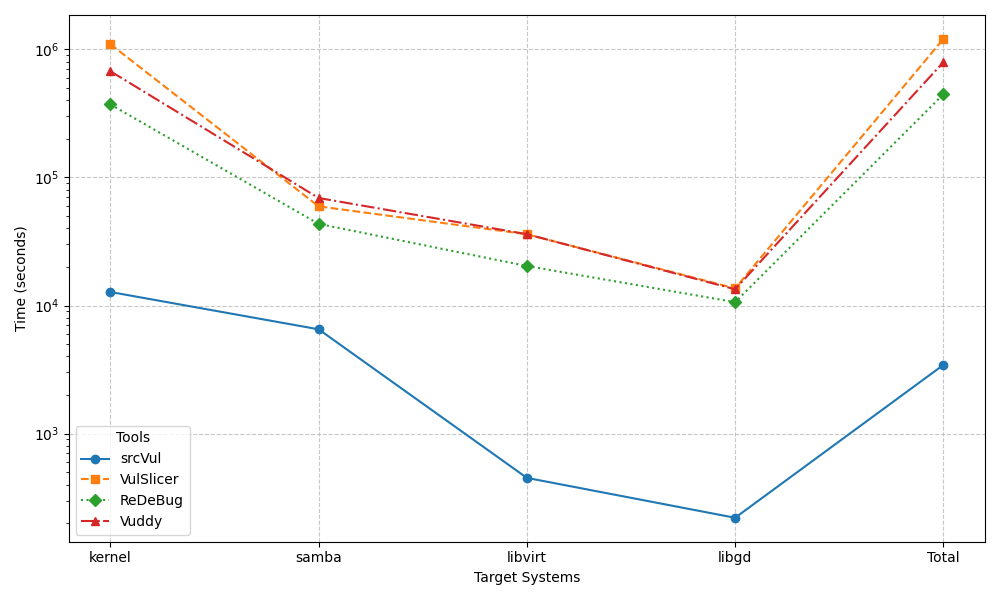}
    \vspace{-2em}
    \caption{Runtime comparison of \textsc{srcVul} and others. }
    \label{fig:archi4}
    \vspace{-1em}
\end{figure}

To answer \textbf{RQ2}, we measured the runtime for each approach across the target systems. The differences in system size led to variability in the number of entities tracked by each method, such as slices for \textsc{srcVul}, function signatures for \textsc{Vuddy}, statements for \textsc{ReDeBug} and \textsc{VulSlicer}. We present the runtime as compared to the number of files, LOC, and CVEs to demonstrate scalability as shown in~\cref{fig:archi4}. \textsc{srcVul} demonstrates lower runtime across all systems, achieving efficient scalability compared to other techniques.

\begin{figure*}
\begin{minipage}{\linewidth}
\begin{lstlisting}[frame=single, basicstyle=\footnotesize\ttfamily, linewidth=\linewidth, breaklines=true, postbreak=\mbox{\textcolor{gray}{$\hookrightarrow$}\space}]
{"CVE_ID": "CVE-2015-7550", "Commit_Hash": "0c47f5e06108c877597de831903d653bcd03a2c2.diff", "Project": "M-Bab__linux-kernel-amdgpu", "Vector_Hash": "72d4709d92231669d063f88c4a2aa4350767d8346f07bb64-17fe154a92d3a83f", "Slicing_Vector": {"Slice Count (SC)": 0.046, "Slice Coverage (SCvg)": 0.181, "Slice Identifier (SI)": 0.041, "Slice Spatial (SS)": 0.82}, "Patch_Content": "-	ret = key_validate(key);-	if (ret == 0) {-		ret = -EOPNOTSUPP;-		if (key->type->read) {-			down_read(&key->sem);+	ret = -EOPNOTSUPP;+	if (key->type->read) {+		down_read(&key->sem);+		ret = key_validate(key);+		if (ret == 0) ret = key->type->read(key, buffer, buflen);-			up_read(&key->sem);-		}+		up_read(&key->sem);"}
\end{lstlisting}
			\vspace{-0.5em}
            \caption{CVE-2015-7550 diff file: improper semaphore use in \texttt{keyctl\_read\_key} (Linux kernel \texttt{security/keys/keyctl.c} pre-4.3.4).}

            \vspace{-0.5em}
            \label{listing:vul}
\end{minipage}
\end{figure*}

Lastly, we evaluated \textbf{RQ3}, which investigates \textsc{srcVul}'s ability to recommend patches after identifying vulnerabilities in the target systems. The recommended patches were manually inspected and evaluated. To illustrate this, the $vs_{vector}$ for \texttt{parent} variable in the target system linux-4.14.76 as shown in~\cref{fig:target} was calculated as: $\langle 0.045, 0.182, 0.045, 0.818 \rangle $. 
We found that this vector is similar to the \texttt{diff} file presented in~\cref{listing:vul}. A closer look at the patch, we can observe that \texttt{down\_read(\&key->sem)} and \texttt{up\_read(\&key->sem)} are used to manage concurrent access to a \texttt{key} object using a semaphore. This prevents conflicts when multiple threads attempt to read or modify the \texttt{key} data. In the target code in~\cref{fig:target}, \texttt{mutex\_lock(\&parent->access)} and \texttt{mutex\_unlock(\&parent->access)} are used to ensure exclusive access to the \texttt{parent} structure while adding an \texttt{entry} to the \texttt{parent->children} list. The mutex protects against concurrent modifications that could cause corruption or unexpected behavior in the list. Both vectors deal with synchronization and resource protection in concurrent environments, indicating similar types of race condition vulnerabilities or data corruption risks. Recognizing this pattern in the patch can highlight potential areas to investigate in the target code.
\section{Related Work}
\label{sec:related}
Several studies have focused on detecting vulnerable code clones using program slicing techniques. \textsc{VulSlicer}~\cite{salimi2022vulslicer} identifies vulnerability-relevant statements (VRS) for known vulnerabilities, slices the target program, and matches the generated slices against a database of vulnerability slices. This approach can detect the first three types of vulnerable code clones. Song et al.~\cite{song2020program} also leverage program slicing to extract vulnerable statements within functions, using a hashed database for vulnerability detection. This method is limited to C/C++ programs and can only identify the first three types of clone vulnerabilities.


Some tools detect code clone vulnerabilities without relying on slicing. \textsc{ReDeBug}~\cite{jang2012redebug} operates at the line level, scanning source code using a fixed-window size to detect vulnerable patterns. \textsc{Vuddy}~\cite{kim2017vuddy} identifies vulnerabilities based on function signatures by normalizing variables, function calls, data types, and parameters. Both \textsc{ReDeBug} and \textsc{Vuddy} can detect clones that exactly match known vulnerabilities but may miss those with different syntax or structure. Other tools, like \textsc{JoanAudit}~\cite{jaudit}, employ slicing techniques specifically for injection vulnerabilities, such as SQL. Thome et al.~\cite{thomasSlvier} introduced a lightweight slicing-based tool for detecting injection vulnerabilities in Java programs. 

Additional static analysis frameworks, such as \textsc{Sflow}~\cite{SFlow}, \textsc{FlowTwist}~\cite{FlowTwist}, \textsc{Andromeda}~\cite{andromeda}, \textsc{Soot}~\cite{soot}, \textsc{LAPSE+}~\cite{LAPSE+}, \textsc{TAJ}~\cite{TAJ}, and \textsc{Movery}~\cite{woo2022movery}, are also designed to detect security vulnerabilities in Java applications. Recent work has explored thin slicing for vulnerability screening. Dashevskyi et al.\protect~\cite{Dashevskyi2019} adapted it from Sridharan et al.\protect~\cite{Sridharan2007} to focus on producer statements that directly contribute to vulnerable computations, reducing slice size while preserving key vulnerability dependencies. However, it struggles with inter-procedural dependencies and may miss vulnerabilities spanning multiple functions and files.

Beyond slicing-based methods, other approaches aim to detect vulnerabilities without slicing but are often limited by different aspects. For example, Livshits et al.~\cite{Livshits} and Shankar et al.\cite{Shankar} proposed methods specific to particular programming languages. \textsc{VulPecker}~\cite{li2016vulpecker} automates vulnerability detection but faces scalability challenges, while \textsc{VCCFinder}~\cite{vcfinder} generates a high rate of FPs, a problem also encountered by \textsc{ReDeBug} and \textsc{Vuddy}. Yamaguchi et al.~\cite{Yamaguchi1} introduced machine learning-based approaches for detecting vulnerabilities in C/C++, providing an alternative to traditional methods.

\section{Conclusion}  
\label{sec:conclusion}  
This paper presents \textsc{srcVul}, a vulnerability detection approach for large-scale systems that combines variable-level program slicing with semantic representation through slicing vectors ($vs_{vectors}$) to identify vulnerable clones, even with syntactic differences. The key idea of \textsc{srcVul} lies in its ability to capture the precise behavior and context of vulnerability-related variables through detailed slice profiles, ensuring improved recall and detection accuracy.

\textsc{srcVul} achieves its effectiveness by focusing on slices derived from specific variables rather than function- or line-level granularity. This variable-level granularity allows it to capture all relevant computations and dependencies, reducing the likelihood of missing vulnerabilities. The generation of detailed slice profiles, including data and control dependencies (e.g., def, use, dvars, ptrs, cfuncs), ensures that subtle and complex vulnerability patterns are accurately identified. Additionally, by leveraging Locality-Sensitive Hashing (LSH) for semantic similarity matching, \textsc{srcVul} can detect semantically similar slices across diverse codebases, overcoming the limitations of exact-match techniques. Furthermore, the approach validates vulnerability matches by analyzing associated patches, ensuring the identified clones are meaningful and reducing false negatives.

Evaluation across various target systems demonstrates \textsc{srcVul}'s efficiency, accuracy, and speed, surpassing existing tools. Additionally, \textsc{srcVul} uniquely explores patch recommendations by leveraging embedded information in CVEs. While this feature represents a promising step toward automated vulnerability remediation, further research is required to enhance its reliability and effectiveness. Future work will focus on expanding the patch recommendation capabilities by refining similarity-matching techniques and exploring machine-learning models to predict more precise patches.

\bibliographystyle{IEEEtran}
\bibliography{references}
\end{document}